\documentclass[twocolumn,reprint,epsf,graphics,psfig,superscriptaddress]{revtex4}
\usepackage{graphicx,epsfig}
\usepackage{times}
\usepackage{graphics,dcolumn,bm,float}
\usepackage{amssymb,amsmath,rotate,color}
\usepackage[normalem]{ulem}

\begin{document}
\title{Electron beam splitting at topological insulator surface states and a proposal for electronic Goos-H\"{a}nchen shift measurement}
\author{Hassan Ghadiri}
\affiliation{Department of Physics, North Tehran Branch, Islamic Azad University, 16511-53311, Tehran, Iran}
\author{Alireza Saffarzadeh}
\altaffiliation{asaffarz@sfu.ca}
\affiliation{Department of Physics, Payame Noor University, P.O.Box 19395-3697 Tehran, Iran} \affiliation{Department of Physics,
Simon Fraser University, Burnaby, British Columbia, Canada V5A
1S6}

\date{\today}

\begin{abstract}
The hexagonal warping effect on transport properties and
Goos-H\"{a}nchen (GH) lateral shift of electrons on the surface of
a topological insulator with a potential barrier is investigated
theoretically. Due to the warped Fermi surface for incident
electron beams, we can expect two propagating transmitted beams
corresponding to the occurrence of double refraction. The
transmitted beams have spin orientations locked to their momenta
so that one of the spin directions rotates compared to the
incident spin direction. Based on a low-energy Hamiltonian near
the Dirac point and considering Gaussian beams, we derive
expressions for calculating lateral shifts in the presence of
warping effect. We study the dependence of transmission
probabilities and GH shifts of transmitted beams on system
parameters in detail by giving an explanation for the appearance
of large peaks in the lateral shifts corresponding to their
transmission peaks. It is shown that the separation between
two transmitted beams through their different GH shifts can be as
large as a few micrometers which is large enough to be observed
experimentally. Finally, we propose a method to measure the GH
shift of electron beams based on the transverse magnetic focusing
technique in which by tuning an applied magnetic field a
detectable resonant path for electrons can be induced.
\end{abstract}
\maketitle

\section{Introduction}
Topological insulators (TIs) are nonmagnetic insulators with
conducting surface states as a consequence of the nontrivial
 topology of their bulk band structure, which in turn results from strong spin-orbit interaction. The surface states contain an
 odd number of spin-helical Dirac cones and are protected against any disturbance that maintains time-reversal symmetry
 \cite{Hasan,Qi,Hasan1}. In the vicinity of the Dirac point, the electron states can be well-described by massless Dirac equation,
 whereas at energies far enough away from the Dirac point, a distortion induced by surface spin-orbit coupling deforms
 the Fermi surface into a hexagonal snowflake shape \cite{Fu-2009,Chen-2009,Kuroda}. This effect,
called hexagonal warping, has been confirmed by angle-resolved
photoemission spectroscopy \cite{Chen-2009}. Since the surface 
states close to the Fermi level play a decisive role in the
electronic properties of two-dimensional (2D) materials, the hexagonal warping can affect transport
properties on the surface of TIs. Therefore, the warping effect and also topologically protection of 
surface states may lead to a variety of interesting properties
which are important from viewpoint of fundamental physics as well
as novel device applications \cite{C. M. Wang,J.
Wang,Pal,An-2012,Rakyta,Siu,Roy,Hassler,Repin,Yu,Akzyanov,Akzyanov1}.

It is well known that the totally reflected light beam from a
dielectric interface undergoes a lateral displacement
 from the position predicted by the geometrical optics. The study of
 this phenomenon which is known as Goos-H\"{a}nchen (GH) shift \cite{Goos}
 has been developed to partial reflections and also transmitting
 configuration \cite{Tamir,Hsue,Li,Broe}. When an electron beam
 is incident on a boundary separating two regions of different
 densities, the reflected/transmitted beam undergoes a GH shift similar to
 a light beam crossing a boundary between materials with different
 optical indices. Accordingly, the GH shift of electrons in condensed
 matter systems \cite{Wilson,X. Chen,X. Chen1,Gruszecki,Yu2} especially
 in Dirac materials \cite{Beenakker,Wu,Zhai,Ghosh,Chen2,Song,Chen3,Sun,
 Ghadiri,Ghadiri1,Qiu,Azarova,Kuai,Jiang,Zheng,Liu,Liu1,Chattopadhyay}
 has been extensively studied.

The GH shift and transverse displacement, called Imbert-Fedorov (IF) shift, of a light beam on the surface of some Dirac materials, such as graphene and Weyl semimetals, have also been investigated \cite{Wu2017,Ye2019}. The results showed that the optical beam shifts provide a possible scheme for direct measurement of the parameters in these materials \cite{Wu2017,Ye2019}. Moreover, it was shown that the electronic IF shift can be utilized to characterize the Weyl semimetals \cite{Jiang}. On the other hand, the results of electronic beam shifts (EBS) suggest a new generation of nanoelectronic devices based on transition metal dichalcogenides \cite{Sun,Ghadiri,Ghadiri1} and Weyl semimetals \cite{Jiang,Zheng}. Therefore, the study of EBS on topological insulators along with the ability of measuring EBS can potentially provide applications in characterizing the parameters of TIs as well as the fabrication of new TI-based nanodevices.

In this paper, we investigate the propagation of electrons through
a square potential barrier on the surface of a TI by considering
the hexagonal warping effect. We show that due to the warped Fermi
surface, an electron wave impinging onto the barrier can have two
transmitted waves, propagating with different momenta and hence in
different directions, much like the double refraction of light in
anisotropic crystals, demonstrating another optics-like property
of electrons. We derive a formula for calculation of GH shifts of
two transmitted beams in the presence of hexagonal warping. We
show that the beams can be separated spatially due to their
different GH shifts, while at the same time they have different
spin orientations due to their different momenta. However, due to
the difficulty in producing well collimated electron beam, the GH
shift in electronic systems has not been measured so far
\cite{Chen4}.

We present a proposal for experimentally measuring the GH shift
based on the transverse magnetic focusing (TMF) technique in which
by applying a transverse magnetic field one can focus the motion
of electrons/holes in the ballistic regime
\cite{Taychatanapat,Milovanovic,Tsoi}. The TMF has been used to study the
shape of Fermi surfaces \cite{Tsoi}, Andreev reflection \cite{Tsoi,Bhandari4}, 
spin-orbit interaction \cite{Rokhinson},  the angle-resolved
transmission probability  in graphene \cite{S. Chen}, imaging
electron trajectories \cite{Bhandari1,Bhandari2,Bhandari3,Bhandari4}, as well as
proposing a method for measuring warping strength in TIs \cite{Yu}.

In this proposal, by applying a transverse magnetic field on
incident region, the impact point and also incident angle of
electrons at the first interface are controlled, similar to the
experiment of Ref. [\onlinecite{S. Chen}] and also the proposal in Ref. [\onlinecite{Yu}]. The variation of
transverse magnetic field applied on the transmission region
induces a resonant conduction path (measured as a voltage peak) by
which the entry point of electrons at the second interface is
determined, and hence, the GH shift can be measured.

The paper is organized as follows. We introduce our model and
formalism for obtaining transmission probability in Sec. II, where
the scattering wave functions in the presence of warping effect and
the transmission properties of incident electron waves are
discussed in details. In Sec. III, we calculate the GH shift of
electron beams and the spatial beam separation is investigated. In
Sec. IV, we present our proposal for measuring the electronic GH
shift based on the TMF phenomenon. We conclude our findings in Sec. V.

\section{Theoretical model and double refraction}

Surface states of topological insulators with a single Dirac cone
are generally described by the Hamiltonian ($\hbar=1$)
\begin{equation}\label{1}
H=\upsilon_{F} {\bf k}\cdot
(\hat{z}\times\boldsymbol{\sigma})+\frac{\lambda}{2}(k_{+}^{3}+k_{-}^{3})\sigma_{z}\ ,
\end{equation}
where ${\bf k}=(k_x,k_y)$ is the wave vector of electron,
$\boldsymbol{\sigma}$ is the Pauli matrix vector, $k_{\pm}=k_{x}
\pm ik_{y}$ and $\hat{z}$ is a unit vector normal to the surface.
$\upsilon_{F}$ and $\lambda$ are the Fermi velocity and warping
parameter, respectively. The linear term in ${\bf k}$ being
similar to that of graphene, with exception that
$\boldsymbol{\sigma}$ represents the real spin of electron and
cubic terms in the Hamiltonian are responsible for hexagonal
warping effect. This Hamiltonian which neglects multi-orbital
structure of surface states is considered as a minimal model
preserving given $C_{3v}$ symmetries \cite{Fu-2009,Akzyanov1}.
Among different topological insulators, Bi$_{2}$Te$_{3}$ is found
to have strong warping effect with Fermi velocity
$\upsilon_{F}=2.55$ eV$\cdot$\AA\ and the hexagonal warping parameter
$\lambda=250$ eV$\cdot${\AA}$^{3}$ that we consider here in the
calculations \cite{Fu-2009,An-2012}. The eigenvalues of Eq.
(\ref{1}) give the upper and lower bands in ${\bf k}$ space as
\begin{equation}\label{2}
E_{\pm}({\bf k})=\pm \sqrt{(\upsilon_{F}k)^{2}+w^{2}({\bf k})}\ ,
\end{equation}
where $w({\bf k})=\lambda k_{x}(k_{x}^{2}-3k_{y}^{2})$. We have
depicted the Dirac cone of fermions (Eq. (2)) and several constant energy
contours (CECs) in Figs. 1(a) and 1(b), respectively. The energy of
CEC is expressed in terms of
$E_{0}=\sqrt{\upsilon_{F}^{3}/\lambda}$ which is the
characteristic energy introduced by hexagonal warping. At $E\ll
E_{0}$ the warping effect is negligible and the CEC exhibits a
circular shape. As the energy exceeds the critical value
$E_{c}=\frac{\sqrt{7}}{6^{\frac{3}{4}}}E_{0}\approx0.69E_{0}$ the
CEC deforms into a hexagonal shape, with inflection points
satisfying the relation $(\frac{\partial k_{x}}{\partial
k_{y}})_{E}=(\frac{\partial^{2} k_{x}}{\partial^{2}
k_{y}})_{E}=0$. With further increasing $E$, the rounded tips of
the hexagon become sharper and the CEC exhibits a snowflake shape.
The eigenspinors of Hamiltonian (1) can be written as
\begin{eqnarray}\label{3}
\chi({\bf k},E_{\pm})\mathrm{e}^{i{\bf k}\cdot{\bf r}}
=\frac{1}{N_{\pm}({\bf k})}\left(\begin{array}{c}
\pm(E_{\pm}({\bf k})+w({\bf k}))\\ \upsilon_{F}(ik_{x}-k_{y})\\
\end{array}\right)\mathrm{e}^{i{\bf k}\cdot{\bf r}}  ,
\end{eqnarray}
where $N_{\pm}({\bf k})$ are the normalization coefficients and 
the subscript $+(-)$ corresponds to the upper (lower) band in Eq. (2). Note
that the interaction between surface states and bulk states can be
ignored since the surface Dirac point on TIs such as
Bi$_{2}$Te$_{3}$ is closer to the bulk valence band than to the
bulk conduction band \cite{Chen-2009}.
\begin{figure}[ht]
\begin{center}
\includegraphics[scale = 0.5]{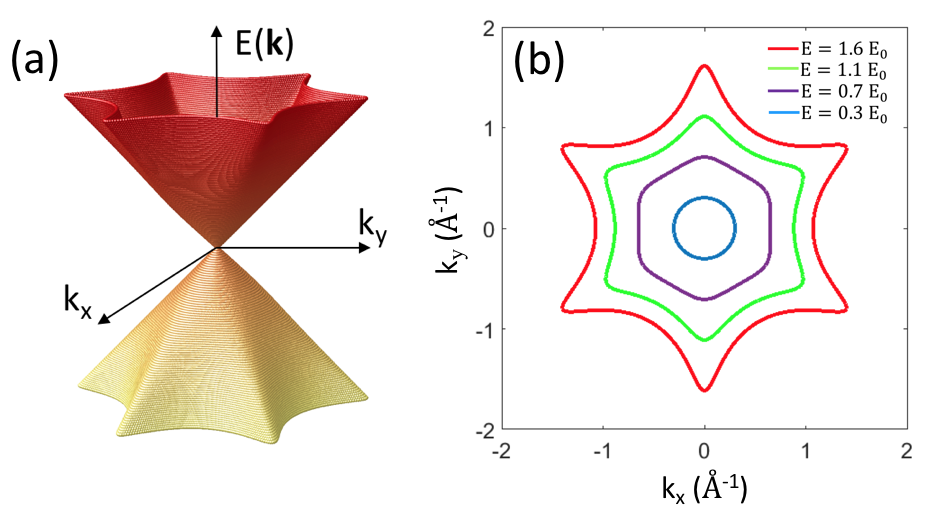}
\caption{(Color online) (a) The Dirac cone of fermions with Fermi
energy $E_{F} = 1.6 E_{0}$ on the surface of Bi$_{2}$Te$_{3}$ by 
including the hexagonal warping effect. (b) The constant-energy
contours of Dirac cone at different energies measured from the
Dirac point.}\label{F1}
\end{center}
\end{figure}

Now, we consider the propagation of electrons
on the surface of TI scattered by a potential barrier $V(y)=V_{0}$
for $0\leqslant y\leqslant d$ and $V(y)=0$ elsewhere (see Fig. 2). Such a
barrier can be produced by a gate electrode deposited on top of
the TI surface. We note that the electron transport in $y$
direction is coherent. Also, due to the translational invariance
in $x$ direction, $k_{x}$ is a good quantum number, and hence, the
Fermi energy $E_{F}$  of electron is conserved in the scattering
process. In zero potential regions for given $E_{F}$ and $k_{x}$,
equation $E_{+}({\bf k})=E_{F}$ is quartic in terms of $k_{y}$ and its
roots determine the $y$ components of electron Fermi momentum. It
gives two real symmetric roots and two imaginary symmetric roots
in the case of $E_{F}<E_{c}$, indicating that an incoming electron
wave with an arbitrary incident angle $\theta$ has one propagating
reflected wave and one propagating transmitted wave (see Fig. 2a)
as it is a normal case for most conventional materials. For
$E_{F}>E_{c}$, the CEC has concave segments and consequently, as
shown in Fig. 2(b), there exists a critical incident angle
$\theta_{c}$ beyond which the equation $E_{+}({\bf k})=E_{F}$ has
four symmetrical real roots. This means that for an incoming
electron wave there exist two propagating reflected waves (double
reflection) and simultaneously two propagating transmitted waves
(double refraction) (see Fig. 2c). When $\theta<\theta_{c}$, similar
to the case of $E_{F}<E_{c}$, single refraction happens as can be
seen in Fig. 2(b). We should note that at $\theta>\theta_{c}$, the
Fermi momenta with bigger absolute values along $y$ axis are
parallel to their corresponding group velocities defined by
$\upsilon_{y}=(\frac{\partial E}{\partial k_{y}})_{k_{x}}$,
therefore  the corresponding states are electron-like. In contrast
the Fermi momenta with smaller absolute values are antiparallel
to their corresponding group velocities indicating hole-like
propagating states. In the barrier region, the roots of $k_{y}$
are obtained from the equation $E_{+}({\bf k})+ V_{0}=E_{F}$
depending on the amounts of $E_{F}, V_{0}$ and incident angle
$\theta$. They can be real, complex, or two real roots and two
imaginary roots.

In order to obtain transmission probability and for the future
purposes, we write down the generic scattering states for given
$E_{F}, V_{0}$ and $\theta=\arctan(|\frac{k_{x}}{k_{y,{1}}}|)$ as
\begin{equation}\label{4}
\psi(y)=\left\{%
\begin{array}{ll}
    \chi(k_{y,1}, E_{F} )e^{ik_{y,1}y}+r_{1}\chi(-k_{y,1}, E_{F} )e^{-ik_{y,1}y}\\
    +r_{2}\chi(k_{y,2}, E_{F} )e^{ik_{y,2} y}, & \hbox{$y \leq 0$,} \\ \\
    \sum_{n=1}^{4} a_{ n}\chi(k_{y,n}^{'}, E_{F}-V_{0} )e^{ik_{y,n}^{\prime}y}, & \hbox{$0\leq y \leq d$,} \\ \\
    t_{1}\chi(k_{y,1}, E_{F} )e^{ik_{y,1}(y-d)}\\ 
    +t_{2}\chi(-k_{y,2}, E_{F} )e^{-ik_{y,2}(y-d)}, & \hbox{$y \geq d$,}
\end{array}%
\right.
\end{equation}
where $\chi(k_{y,1}, E_{F} )$ is the incident state with Fermi
momentum $k_{y,1}$, $r_{1}(r_{2})$ is the reflection amplitude
corresponding to the Fermi momentum $-k_{y,1}(k_{y,2})$, and
$t_{1}(t_{2})$ is the transmission amplitude corresponding to the
Fermi momentum $k_{y,1}(-k_{y,2})$, while $a_{ n}$ is the
scattering amplitude corresponding to the momentum $k_{y,n}^{'}$
in the barrier region. As can be seen in Fig. 2, at $E_{F}>E_{c}$
and $\theta_{c}<\theta<\frac{\pi}{3}$ $(\theta>\frac{\pi}{3})$,
$k_{y,1}$ is a positive electron-like (negative hole-like)
momentum and $k_{y,2}$ is a positive hole-like (negative
electron-like) momentum, while for $\theta<\theta_{c}$, $k_{y,1}$
is a positive real root and $k_{y,2}$ is a negative imaginary
root. In the case of $E_{F}<E_{c}$, however, for every incident
angle $\theta$, $k_{y,1}$ is a positive real root and $k_{y,2}$ is
a negative imaginary root.
\begin{figure}[ht]
\begin{center}
\includegraphics[scale = 0.55]{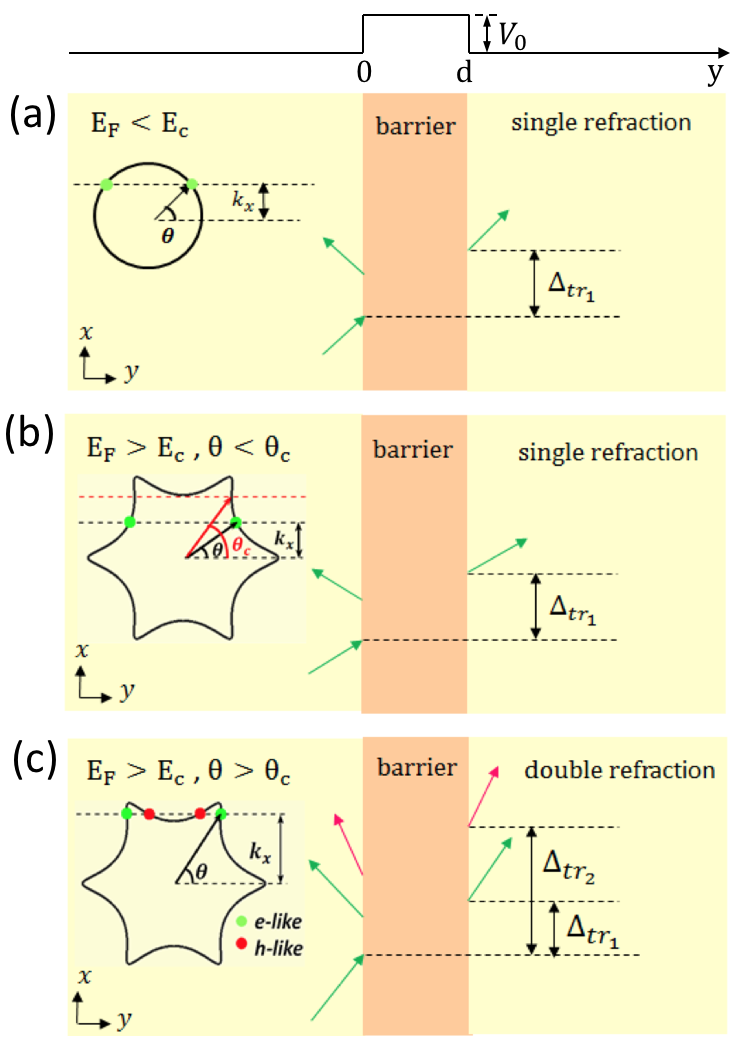}
\caption{(Color online) Scattering processes when an incident
electron is reflected and transmitted from the barrier with width
$d$ and height $V_{0}$ as shown on top of (a) in the cases of
(a,b) single and (c) double refractions. The insets show the CEC at
(a) $E_{F}<E_{c}$, (b)  $E_{F}>E_{c}$,  $\theta<\theta_{c}$ and
(c) $E_{F}>E_{c}$,  $\theta>\theta_{c}$. The green (red) circles
on CECs indicate propagating electron (hole)-like states. Also,
the GH shift $\Delta_{tr_{1}}$ of beam 1 with the same momentum as
that of the incident beam and GH shift $\Delta_{tr_{2}}$ of beam 2
with different momentum compared to the incident beam are
shown.}\label{F2}
\end{center}
\end{figure}

The eigenvalue equation $(H+V(y))\psi(y)=E\psi(y)$ corresponding to the Hamiltonian (1) is a 
second-order partial differential equation with respect to $y$, due to the warping effect. Therefore, by 
applying boundary conditions of continuity of $\psi(y)$ and its
first derivative with respect to $y$ at the two interfaces $y=0$
and $y=d$, the reflection and transmission amplitudes, and also,
the scattering amplitude $a_{n}$ can be determined. The
transmission probability which is defined as the ratio of
$y$-component of the probability current density of the
transmitted waves and that of the incident wave can be expressed
in terms of the transmission amplitudes and the $y$-component of the
corresponding group velocities as
$T=T_{1}+T_{2}=|t_{1}|^{2}+\frac{\upsilon
_{y}(-k_{y,2},E_{F})}{\upsilon _{y}(k_{y,1},E_{F})}|t_{2}|^{2}$
in the case of double refraction and $T=T_{1}=|t_{1}|^{2}$ in the case
of single refraction \cite{M. Wu,Arabikhah}. In these relations
$T$ denotes the total transmission probability, while $T_{1}$
($T_{2}$) represents the transmission probability of transmitted
wave with the same (different) momentum as (from) the incident wave
momentum.

We have shown in Fig. 3 the contour plot of the transmission probability as a function of incident angle and the potential barrier height for two different Fermi energies $E_{F} < E_{c}$ and $E_{F} > E_{c}$. At $E_{F} < E_{c}$, there is only one propagating transmitted wave whose probability at the typical Fermi energy of 0.15 eV and the barrier width $d=500$\AA\ is shown in Fig. 3(a). As can be seen, there is a region with two boundaries inside which the total internal reflection (TIR) takes place. The boundaries represent the geometrical locations of a critical angle $\theta_{TIR}=\theta_{TIR}(V_{0})$ such that when the incident angle reaches $\theta_{TIR}$, all four waves inside the barrier region become evanescent. Therefore, at sufficiently wide barrier, TIR begins. To obtain an expression for $\theta_{TIR}$, first we consider equation $E_{+}({\bf k})+V_{0}=E_{F}$ which is quadratic in terms of $k_{y}^{2}$.  By solving the discriminant of this equation, $k_x$ value corresponding to $\theta_{TIR}$ can be obtained. Replacing $k_x$ in equation $E_{+}{\bf (k)}=E_{F}$, we obtain the corresponding $k_y$ of $\theta_{TIR}$, resulting $\theta_{TIR}=\arctan{(k_x/k_y)}$. The obtained analytical expression for $\theta_{TIR}$ is complicated. Therefore, we have not presented the resulting expression here. Instead, we explain below how $\theta_{TIR}$ occurs and behaves as a function of $V_{0}$. At $E_{F} =0.15$ eV and the typical value $V_{0}=0.05$ eV, the CECs in the incident and barrier regions are shown in Fig. 4(a). The size of CEC in the barrier region (determined by $|V_{0}-E_{F}|$) is smaller than the size of CEC in the incident region. The line $kx=cte.$, corresponding to the incident angle $\theta$ represents the conservation of $k_{x}$ in the electron scattering process. As shown in Fig. 4(a), at a small incident angle $\theta$, the line $k_{x}=cte.$ intersects the CEC of the barrier region at two points, representing two real wave numbers of propagating electrons inside the barrier. With increasing $\theta$, the corresponding line $k_{x}=cte.$ moves upwards until it touches the CEC of the barrier region at a single point, indicating that the incident angle $\theta$ reaches the critical angle $\theta_{TIR}$. When $\theta$ exceeds $\theta_{TIR}$, the line of $k_{x} = cte.$ can no longer intersects the CEC of the barrier region, meaning that all $k'_{y,n}$ are complex and consequently, TIR begins. With increasing $V_{0}$ the size of CEC in the barrier region decreases, and hence, the critical angle becomes smaller. As $V_{0}$ approaches $E_{F}$, the critical angle at which all $k'_{y,n}$ become complex approaches zero. However, due to the Klein tunneling effect (see discussion below) which prohibits backscattering near the normal incident angle, TIR cannot start at $\theta=0^{\circ}$ (see Fig. 3(a)) \cite{Article3}. When $V_{0}$ exceeds $E_{F}$, the size of CEC in the barrier region increases again as well as the critical angle, until $|V_{0}-E_{F}|$ reaches $E_{F}$ . From now on, since the size of CEC in the barrier region becomes larger than that of the CEC in the incident region, TIR does not form at any incident angle.

As can be seen in Fig. 3(a), the barrier remains perfectly transparent at incident angles close to the normal incidence $\theta=0^{\circ}$, regardless of the amount of $V_{0}$. This process is known as Klein tunneling \cite{Klein} and originates from spin conservation \cite{Katsnelson}, since nonmagnetic barriers cannot change the spin direction of incident electrons in a scattering process. On the other hand, at almost normal incidence, the spin states of incident wave  and propagating reflected wave are orthogonal (this can be easily deduced from Eq. (3)). Therefore, backscattering is forbidden and electrons can transmit perfectly. By increasing $\theta$, the spin states of incident and propagating reflected waves are no longer orthogonal, and hence, the electron reflection is allowed. Now, we consider the case of oblique incidence. When a wave impinges on the barrier at a given $\theta\ne 0$, a part of wave transmits into the barrier and is multiply reflected at the two interfaces $y=0$ and $y=d$, therefore interference happens. As $V_{0}$ increases from zero, the size of CEC in the barrier region changes, and hence, the acquired phase $k’_{y}d$ of propagating waves inside the barrier region will change, causing oscillations in transmission probability. Whenever the waves interfere constructively, the Fabry-P\'{e}rot resonances with $T(\theta\ne 0)=1$ appears. As $V_{0}$ reaches a value at which the line $k_{x}=cte.$ becomes tangent to the CEC of barrier region (see Fig. 4(a)), the TIR begins. It continues until the line $k_{x}=cte.$ again becomes tangent to the CEC, but this time at an amount of 
$V_{0}>E_{F}$. With further increasing $V_{0}$, oscillations appear again in the transmission spectrum. Moreover, similar to the Schr\"{o}dinger-type electrons, when the potential height is less than Fermi energy (corresponds to a n-n$'$-n junction), the incident electrons generally pass through the potential barrier with larger transmission probabilities, compared to the case of $V_{0}>E_{F}$ (corresponds to a n-p-n junction). Also, we should mention that for $\theta>80^{\circ}$ the group velocity of electrons $v_{y}=(\frac{\partial E}{\partial k_{y}})_{k_{x}}=-v_{x}(\frac{\partial k_{x}}{\partial k_{y}})_{E}\simeq 0$ (see CEC in incident region in Fig. 4(a)). Therefore, the electron transmission becomes very small at $V_{0}=0$. However, at some $V_{0}$ values, $T$ can be considerably magnified, due to the interference effect. It is important to point out that for validity of minimal continuum model described by Hamiltonian (1), both $E_{F}$ and $|V_{0}-E_{F}|$ must be considered less than 0.4 eV \cite{Chen-2009,An-2012}. However, in Fig. 3(a), we terminate $V_{0}$ at 0.4 eV as at larger values no more features can be observed.

\begin{figure}[ht]
\begin{center}
\includegraphics[scale = 0.53]{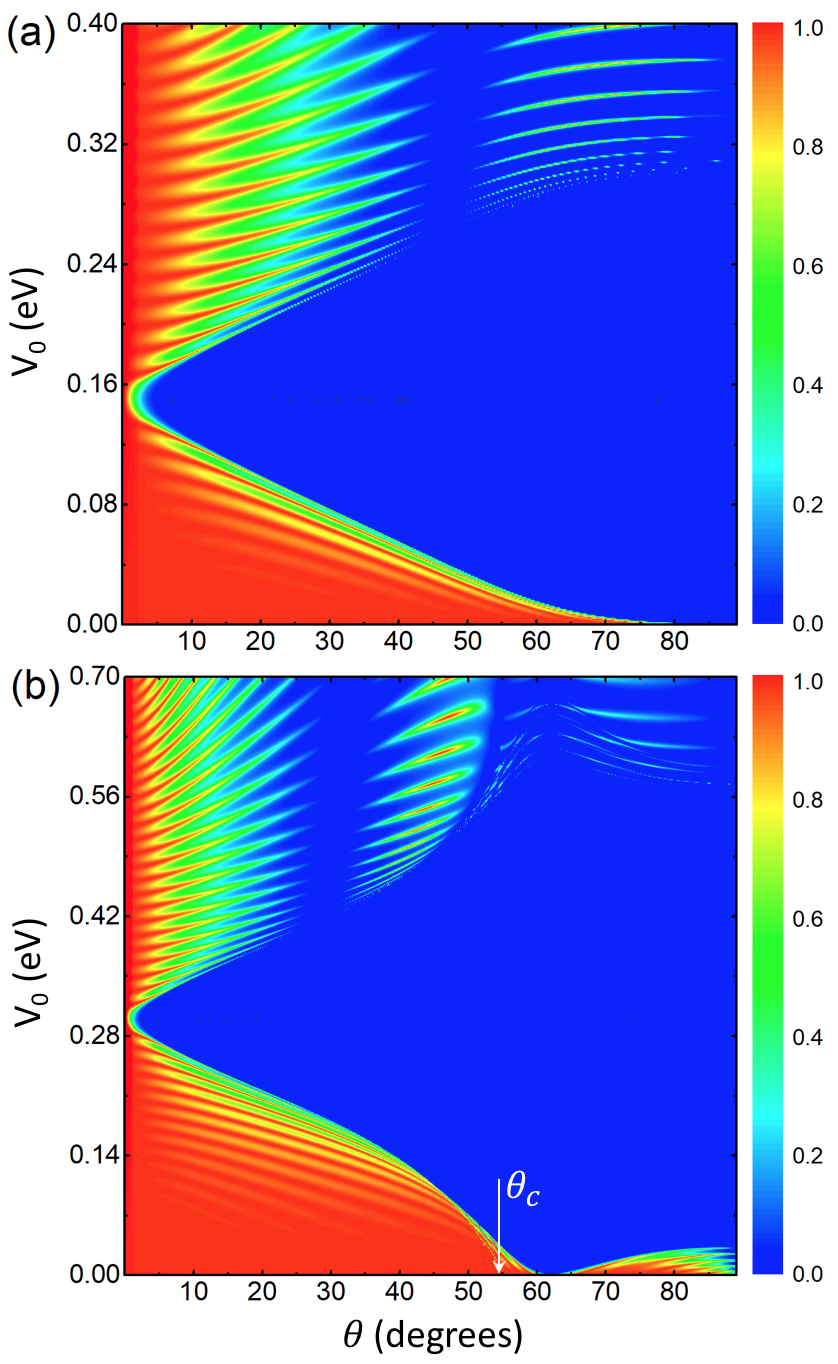}
\caption{(Color online) Calculated transmission probability
$T_{1}$ versus $\theta$ and $V_{0}$ for incident electrons on a
potential barrier of width $d=500$ \AA\,at Fermi energy (a)
$E_{F}=0.15$ eV$<E_{c}$ and (b) $E_{F}=0.3$ eV$>E_{c}$. ($E_{c}\simeq 0.18$ eV)}\label{F3}
\end{center}
\end{figure}

\begin{figure}[ht]
\begin{center}
\includegraphics[scale = 0.45]{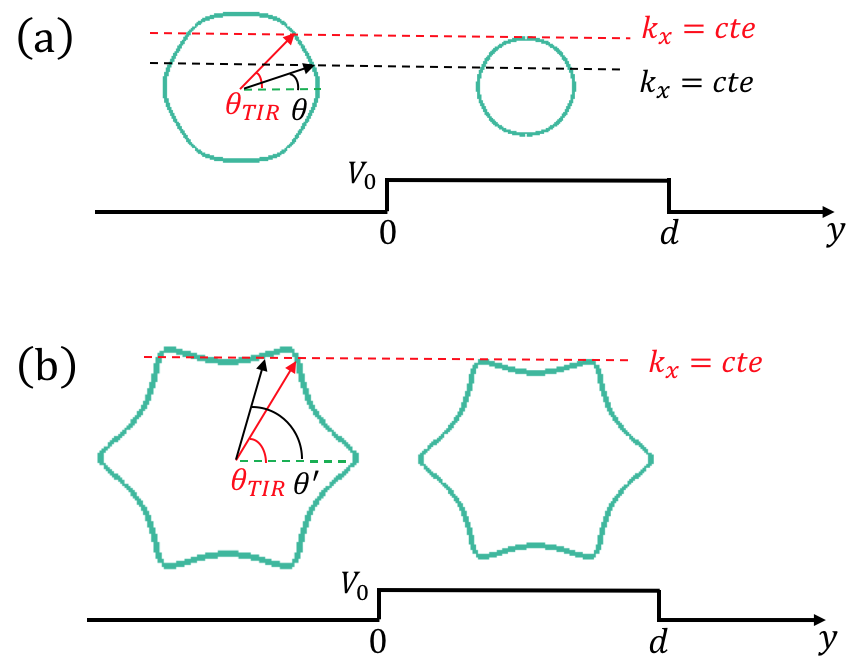}
\caption{(Color online) (a) Solid curves represent CECs in the incident 
and barrier regions for $E_{F}=0.15$ eV and $|V_{0}-E_{F}|=0.1$ eV. The black 
(red) dashed line shows the conservation of transverse momentum $k_{x}$ in the scattering process 
at incident angle $\theta(\theta_{TIR})$. (b) The CEC in the incident region corresponds to 
$E_{F}=0.3$ eV. The value of $V_{0}$ is chosen such that the maximum of $k_x$ on the 
CEC of the barrier region is bigger than the $k_x$ value on CEC of the incident region at 
$\theta=\pi/2$, i.e. $\theta_{TIR}>\theta_{c}$. Due to the warped CEC in the incident region, TIR is confined in the interval of
$\theta_{TIR}<\theta<\theta'$.}\label{F4}
\end{center}
\end{figure}

\begin{figure}[ht]
\begin{center}
\includegraphics[scale = 0.53]{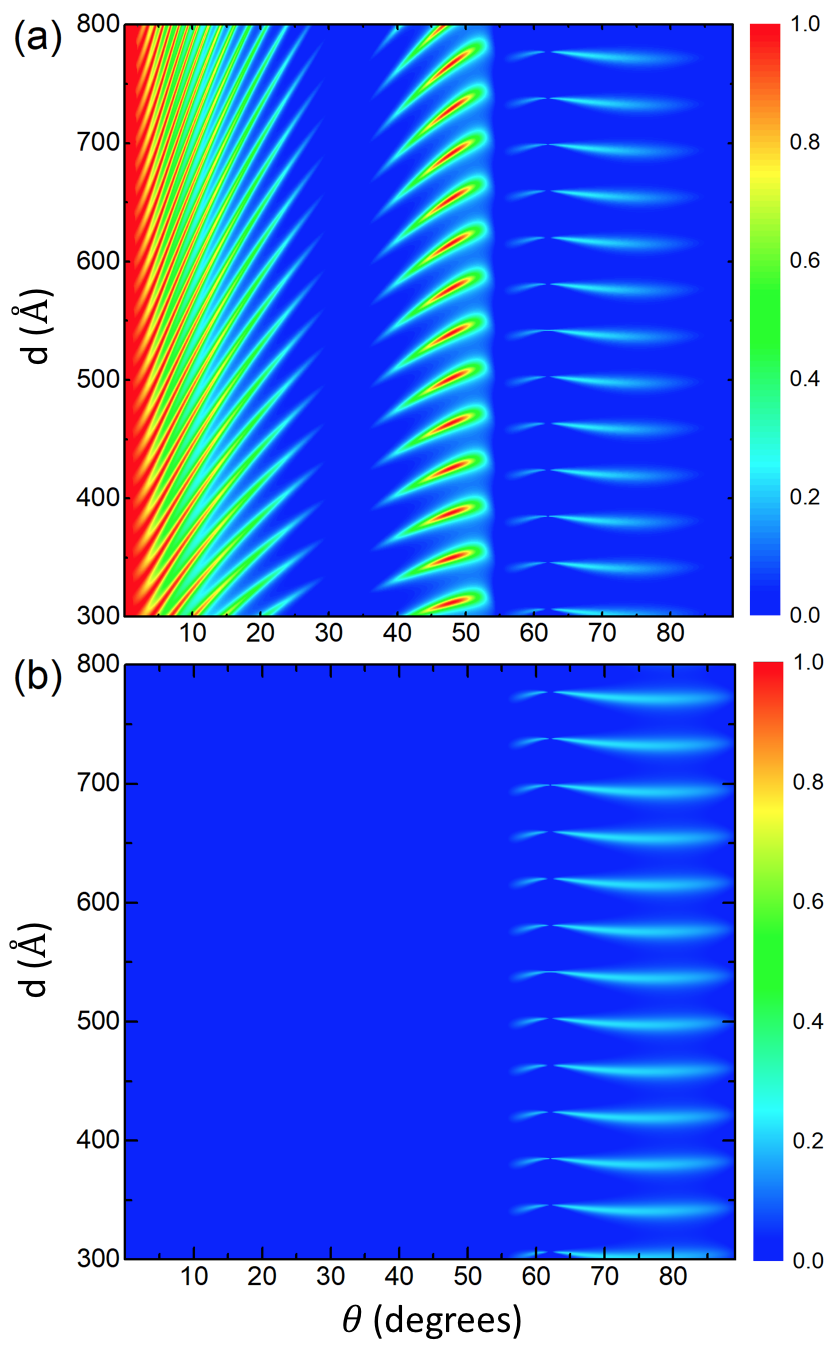}
\caption{(Color online) Calculated transmission probabilities (a)
$T_{1}$ and (b) $T_{2}$ versus $\theta$ and $d$ for incident
electrons with Fermi energy $E_{F}=0.3$ eV on a potential barrier
of $V_{0}=0.7$ eV.}\label{F5}
\end{center}
\end{figure}

The CEC of zero-potential regions is warped at $E_{F}>E_{c}$, and hence, two transmitted waves can propagate. The transmission probability $T_{1}$ at $E_{F}=0.3$ eV and with the same barrier width value as that in Fig. 3(a) is depicted in Fig. 3(b).  First, we consider a given $V_{0}$ at which the size of CEC in the barrier region is small enough compared to the CEC of the incident region. Similar to Fig. 4(a), as $\theta$ increases from zero, the corresponding line $k_{x}=cte.$ moves upwards and consequently, the acquired phase $k'_{y}d$ of propagating electron waves inside the barrier region will change, resulting oscillations in $T_{1}$ and also Fabry-P\'{e}rot resonances in the constructive interference. When $\theta$ reaches $\theta_{TIR}$, the line $k_{x}=cte.$ touches the CEC of barrier region at a single point, that is the start of TIR. The TIR extends to $\pi/2$ because for $\theta>\theta_{TIR}$ the constant $k_{x}$ line can no longer intersect the CEC of the barrier region. Now, we consider an amount of $V_{0}$ at which the size of CEC in the barrier region is close enough to the CEC of incident region, so that the maximum of $k_{x}$ in the barrier region is larger than $ k_{x}$ at $\theta=\pi/2$  in the incident region (see Fig. 4(b)). In this case, when $\theta$ exceeds $\theta_{TIR}$, the corresponding constant $k_{x}$ line will start to intersect the CEC of the barrier region one more time at an angle $\theta'>\pi/3$. Therefore, the TIR terminates at $\theta'$ and will not extend to $\pi/2$. This happens at $V_{0}<0.03$ eV and 0.57 eV$<V_{0}<$0.6 eV in Fig. 3(b). At almost normal incident angle, the Klein tunneling happens, regardless of $V_{0}$ value, similar to Fig. 3(a). However, at a given oblique incident angle $\theta\ne 0$ when $V_{0}$ varies outside the TIR region, the size of CEC in the barrier region changes, and hence, the acquired phase $k’_{y}d$ of the propagating waves inside the barrier region may also change, causing  an oscillatory behavior in $T_{1}$, similarly to the behavior of Fig. 3(a). Since $(\partial k_{x}/\partial k_{y})_{E}$ approaches zero as $\theta$ reaches $62^{\circ}$ (see the CEC in the incident region in Fig. 4(b)), $v_{y}$ and consequently $T_{1}$ are very small, independent of $V_{0}$ values. On the other hand, the transmission probability $T_{2}$ (not shown here) is zero at $\theta<\theta_{c}=54.7^{\circ}$, while it has the same features as $T_{1}$ at $\theta>\theta_c$, regardless of $V_{0}$ values.

The transmission probabilities $T_{1}$ and $T_{2}$ for the two transmitted waves 1 and 2 as functions of incident angle $\theta$ and the barrier width $d$, at $E_{F}=0.3$ eV and $V_{0}=0.7$ eV are shown in Figs. 5(a) and 5(b), respectively. Since $|V_{0}-E_{F}|>E_{F}$, the TIR does not occur and for a given $d$ value at almost normal incident angle, the Klein tunneling with perfect transmission probability for $T_{1}$ happens, as shown in Fig. 5(a). As $\theta$ is increased from zero, the acquired phase $k’_{y}d$ of the propagating waves inside the barrier region changes. This causes oscillations in $T_{1}$, emerging Fabry-P\'{e}rot resonances when constructive interference takes place. If $d$ varies at a fixed oblique incident angle, the acquired phase $k’_{y}d$ will change so that $T_{1}$ again exhibits oscillations and Fabry-P\'{e}rot resonances can emerge. Here, $\theta_{c}=54.7^{\circ}$ is the same as that in Fig. 3(b) because $\theta_{c}$ depends only on $E_{F}$. When $\theta$ exceeds $\theta_{c}$, $T_{2}$ in Fig. 5 (b) takes non-zero values and shows oscillations due to the change of acquired phase $k’_{y}d$, with varying $\theta$ or $d$. In the vicinity of  $\theta=62^{\circ}$, both $T_{1}$ and $T_{2}$ are very small for the same reason explained in Fig. 3(b).

\begin{figure}[ht]
\begin{center}
\includegraphics[scale = 0.73]{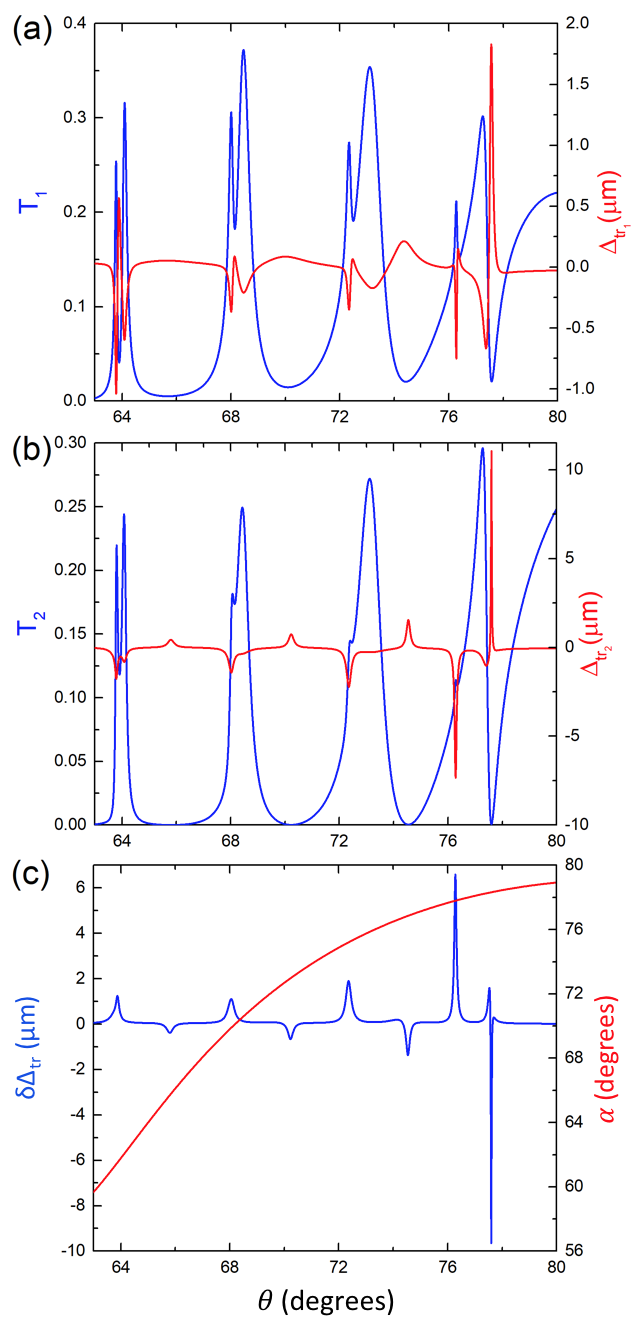}
\caption{(Color online) (a,b) Calculated transmission
probabilities $T_{1,2}$ (blue curves) and the Gh shifts
$\Delta_{tr_{1,2}}$ (red curves) in transmission as a function of
$\theta$. (c) The spatial separation $\delta\Delta_{tr}$ (blue
curve) and the angle difference $\alpha$ (red curve) between spin
orientation of transmitted beams. The parameters are
$E_{F}=0.35$ eV, $V_{0}=0.72$ eV, and $d=$490 \AA.}\label{F6}
\end{center}
\end{figure}

\section{Goos-H\"{a}nchen shift and beam splitting}
The GH shift for a plane wave of electrons is not detectable due to its infinite
spatial width. Therefore, to calculate the GH lateral shift, we consider a beam of electrons
instead of a plane wave. We model an incident electron beam by using a
Gaussian wave packet of surface states as
\begin{equation}\label{5}
    \psi_{in}({\bf r})=\int_{-\infty}^{+\infty} dk_{x}f(k_{x}-k_{x_{0}})\chi(k_{y_{,1}}(k_{x}),E_{F})e^{i(k_{x}x+k_{y_{,1}}(k_{x})y)},
\end{equation}
where
$f(k_{x}-k_{x_{0}})=(\sqrt{2\pi}\Delta_{k_{x}})^{-1}e^{-(k_{x}-k_{x_{0}})^{2}/2\Delta_{k_{x}}^{2}}$
shows the Gaussian angular distribution of width $\Delta_{k_{x}}$
around central incident angel
$\theta_{0}=\arctan(|\frac{k_{x_{0}}}{k_{y,{1}}}|)$. Analogously,
the wave functions of transmitted electron beams can be written as
\begin{eqnarray}\label{6}
    \psi_{tr_{1}}({\bf r})&=&\int_{-\infty}^{+\infty} dk_{x}f(k_{x}-k_{x_{0}})t_{1}(k_{x})\nonumber\\
   && \times\chi(k_{y_{,1}}(k_{x}),E_{F})e^{i(k_{x}x+ k_{y_{,1}}(k_{x}) (y-d))},
\end{eqnarray}
and
\begin{eqnarray}\label{7}
    \psi_{tr_{2}}({\bf r})&=&\int_{-\infty}^{+\infty} dk_{x}f(k_{x}-k_{x_{0}})t_{2}(k_{x})\nonumber\\
    && \times\chi(-k_{y_{,2}}(k_{x}),E_{F})e^{i(k_{x}x- k_{y_{,2}}(k_{x}) (y-d))}.
\end{eqnarray}

For the well collimated electron beams, $f(k_{x}-k_{x_{0}})$ is
sharp around $k_{x_{0}}$ such that the spinor components
$\chi^{\pm}=\mid\chi^{\pm}\mid e^{i\varphi^{\pm}}$ can be
converted into an exponential form and approximated by keeping the
first two terms of the Taylor expansion of its exponent around
$k_{x_{0}}$ as
\begin{eqnarray}\label{8}
    \chi^{\pm}(k_{y_{,1}}(k_{x}))&=&\exp[\ln\chi^{\pm}(k_{y_{,1}}(k_{x}))]\nonumber\\
    &\simeq&\chi^{\pm}(k_{y_{,1}}(k_{x_{0}}))\exp\{[\frac{|\dot{\chi}^{\pm}(k_{y_{,1}}(k_{x_{0}}))|}{|\chi^{\pm}(k_{y_{,1}}(k_{x_{0}}))|}\nonumber\\
    &&+i\dot{\varphi}^{\pm}(k_{y_{,1}}(k_{x_{0}}))](k_{x}-k_{x_{0}})\},
\end{eqnarray}
where
$\dot{\varphi}^{\pm}(k_{y_{,1}}(k_{x_{0}}))(|\dot{\chi}^{\pm}(k_{y_{,1}}(k_{x_{0}}))|)$
denotes derivative of
$\varphi^{\pm}(k_{y_{,1}}(k_{x}))(|\chi^{\pm}(k_{y_{,1}}(k_{x}))|)$
with respect to $k_{x}$, evaluated at $k_{x}=k_{x_{0}}$.
Substituting Eq. (8) into Eq. (5), using the approximation
$k_{y_{,1}}(k_{x})\simeq
k_{y_{,1}}(k_{x_{0}})+\dot{k}_{y_{,1}}(k_{x_{0}})(k_{x}-k_{x_{0}})$
and then evaluating the integral we obtain the spatial form of the
components of the incident beam as
\begin{eqnarray}\label{9}
    \psi_{in}^{\pm}({\bf r})&=&\chi^{\pm}(k_{y_{,1}}(k_{x_{0}}))\nonumber\\
    &&\times e^{-[x+\dot{\varphi}^{\pm}(k_{y_{,1}}(k_{x_{0}}))+\dot{k}_{y_{,1}}(k_{x_{0}})y]^{2}\Delta_{k_{x}}^{2}/2}\nonumber\\
    &&\times e^{\gamma^{\pm^{2}}/2\Delta_{k_{x}}^{2}} e^{i\gamma^{\pm} \dot{\varphi}^{\pm}(k_{y_{,1}}(k_{x_{0}}))} \nonumber\\
    &&\times e^{i[(k_{y_{,1}}(k_{x_{0}})+\gamma^{\pm}\dot{k}_{y_{,1}}(k_{x_{0}}))y+(k_{x_{0}}+\gamma^{\pm})x]},
\end{eqnarray}
where
$\gamma^{\pm}=\frac{\Delta_{k_{x}}^{2}|\dot{\chi}^{\pm}(k_{y_{,1}}(k_{x_{0}}))|}{|\chi^{\pm}(k_{y_{,1}}(k_{x_{0}}))|}$.
As can be seen from the second factor in Eq. (9), the incident beam has a
Gaussian shape and the peak location of its upper and lower
components at the interface $y=0$ is given by
$x_{in}^{\pm}=-\dot{\varphi}^{\pm}(k_{y_{,1}}(k_{x_{0}}))$.
Therefore, the average location of incident beam at the interface
$y=0$ can be expressed as
\begin{eqnarray}\label{10}
    \bar{x}_{in}&=&-\dot{\varphi}^{+}(k_{y_{,1}}(k_{x_{0}}))|\chi^{+}(k_{y_{,1}}(k_{x_{0}}))|^{2}\nonumber\\
    &&-\dot{\varphi}^{-}(k_{y_{,1}}(k_{x_{0}}))|\chi^{-}(k_{y_{,1}}(k_{x_{0}}))|^{2}.
\end{eqnarray}
It is worth mentioning that the last factor in Eq. (9) shows that
the propagation direction of incident-beam components deviates
from the central angle $\theta_{0}$ by the amount of
$\delta^{\pm}\approx\tan\delta^{\pm}\approx\frac{\gamma^{\pm}}{k_{y_{,1}}(k_{x_{0}})}$.
This deflection is due to the warping effect as in the absence of
warping $|\chi^{\pm}|$ is constant and $\delta^{\pm}=0$. Moreover,
the third factor in Eq. (9) reveals that the magnitude of incident
beam is adjusted by warping as well.

By comparing Eqs. (6) and (7) with Eq. (5) we can write an
expression for the transmitted beam components, similar to Eq.
(9), by the substitutions $\chi^{\pm}\mapsto\chi^{\pm}t_{1(2)}$,
$\varphi^{\pm}\mapsto\varphi^{\pm}+\varphi_{t_{1(2)}}$and
$|\chi^{\pm}|\mapsto|\chi^{\pm}||t_{1(2)}|$ in Eq. (9) where
$\varphi_{t_{1(2)}}$ represent the phase of transmission amplitude
$t_{1(2)}$. Therefore, the transmitted beams find also Gaussian
shapes just like incident beam. The average locations of the
transmitted beams at the interface $y=d$ read as
\begin{eqnarray}\label{11}
    \bar{x}_{tr_{1}}&=&-\dot{\varphi}^{+}(k_{y_{,1}}(k_{x_{0}}))|\chi^{+}(k_{y_{,1}}(k_{x_{0}}))|^{2}-\dot{\varphi_{t_{1}}}(k_{x_{0}})\nonumber\\
    &&-\dot{\varphi}^{-}(k_{y_{,1}}(k_{x_{0}}))|\chi^{-}(k_{y_{,1}}(k_{x_{0}}))|^{2},
\end{eqnarray}
and
\begin{eqnarray}\label{12}
    \bar{x}_{tr_{2}}&=&-\dot{\varphi}^{+}(-k_{y_{,2}}(k_{x_{0}}))|\chi^{+}(-k_{y_{,2}}(k_{x_{0}}))|^{2}-\dot{\varphi_{t_{2}}}(k_{x_{0}})\nonumber\\
    &&-\dot{\varphi}^{-}(-k_{y_{,2}}(k_{x_{0}}))|\chi^{-}(-k_{y_{,2}}(k_{x_{0}}))|^{2}.
\end{eqnarray}

\begin{figure}[ht]
\begin{center}
\includegraphics[scale = 0.75]{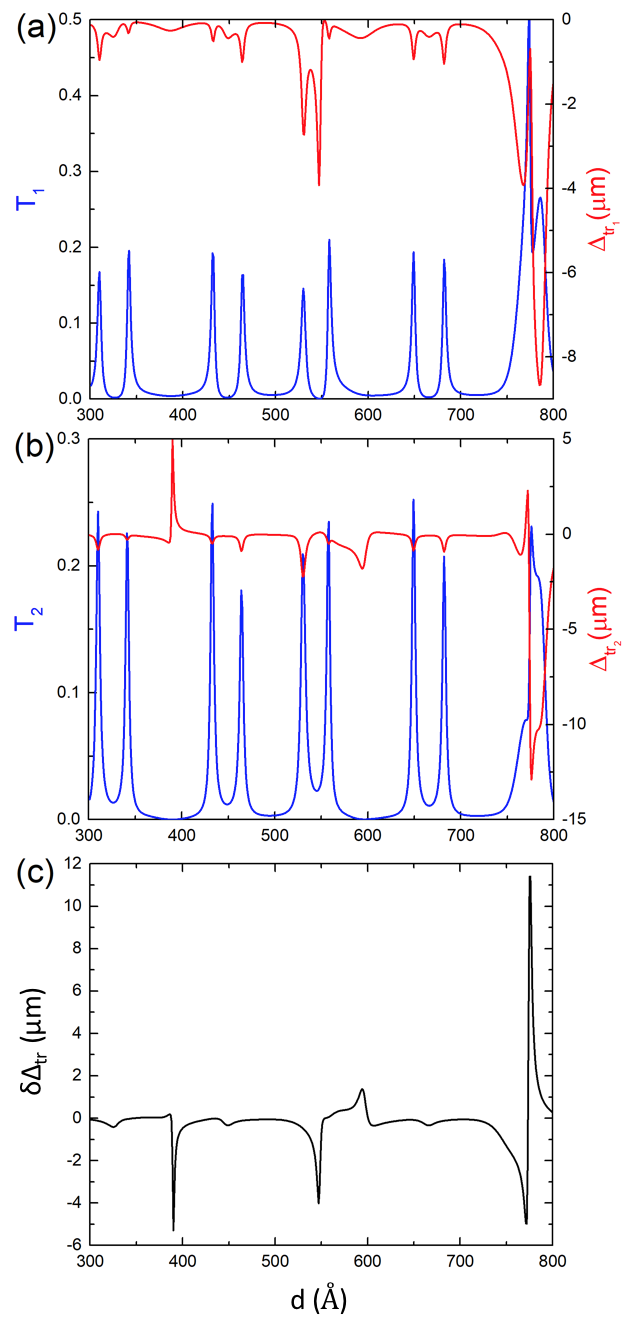}
\caption{(Color online) (a,b) Calculated transmission
probabilities $T_{1,2}$ (blue curves) and the GH shifts
$\Delta_{tr_{1,2}}$ (red curves) in transmission as a function of
barrier width $d$. (c) The spatial separation $\delta\Delta_{tr}$
between transmitted beams. The parameters are $E_{F}=0.3$ eV,
$V_{0}=0.65$ eV, and $\theta=59^{\circ}$.}\label{F7}
\end{center}
\end{figure}

The GH lateral shift $\Delta_{tr}$ is defined as the displacement
of the peak of transmitted beam at the interface $y=d$
relative to the peak of incident beam at the interface $y=0$
\cite{Broe} (see Fig. 2) which is different from classical
prediction of electron optics, i.e., Snell's shift
$d\tan\theta^{\prime}$, where $\theta^{\prime}$ is the refraction
angle. Therefore, the GH shift of the transmitted beam 1(2) with
the same (different) momentum as (from) that of the incident beam can be
obtained from Eqs. (10)-(12) as
\begin{equation}\label{13}
    \Delta_{tr_{1}}=-\dot{\varphi_{t_{1}}}(k_{x_{0}}),
\end{equation}
and
\begin{eqnarray}\label{14}
    \Delta_{tr_{2}}&=&-\dot{\varphi_{t_{2}}}(k_{x_{0}})-\dot{\varphi}^{-}(-k_{y_{,2}}(k_{x_{0}}))|\chi^{-}(-k_{y_{,2}}(k_{x_{0}}))|^{2}\nonumber\\
    &&+\dot{\varphi}^{-}(k_{y_{,1}}(k_{x_{0}}))|\chi^{-}(k_{y_{,1}}(k_{x_{0}}))|^{2}.
\end{eqnarray}

In deriving Eq. (14) we have used
$\dot{\varphi}^{+}(k_{y_{,1}}(k_{x_{0}}))=\dot{\varphi}^{+}(-k_{y_{,2}}(k_{x_{0}}))=0$
because in the case of double refraction upper spinor components in
zero potential regions are real. The spatial splitting between the
two beams will occur when they have different GH shifts. In this case, the
spatial separation between the two beams is given by
$\delta\Delta_{tr}=\Delta_{tr_{1}}-\Delta_{tr_{2}}$.

The electron spin orientation can be obtained from Eq. (3) as
$\mathbf{s}=<\boldsymbol{\sigma}>=E_{+}^{-1}(-\upsilon_{F}k_{y},
\upsilon_{F}k_{x}, w({\bf k}))$ indicating the spin-momentum
locking of surface electrons in TIs, due to the spin-orbit
coupling. Consequently, the spin direction of the transmitted beam
2 is rotated relative to the spin direction of both transmitted
beam 1 and incident beam by the amount of
$\alpha=\arccos(\mathbf{s}_{1}\cdot\mathbf{s}_{2})$ where
$\mathbf{s}_{1}$ and $\mathbf{s}_{2}$ are spin orientations of
transmitted beams 1 and 2, respectively.

Due to the warping term, $\Delta_{tr_{1(2)}}$ cannot be derived in
compact analytical expressions. Therefore, these quantities are
calculated numerically using Eqs. (13) and (14). Typical results
for GH shifts and the corresponding transmission probabilities are
shown in Figs. (6) and (7). The parameters are chosen to avoid TIR
and that two transmitted beams propagate. Figs. 6(a) and (b) show
the transmission probabilities and the corresponding GH values of
the two transmitted beams 1 and 2 in terms of incident angle. Due
to the interference effect, $T_{1,2}$ show an oscillatory behavior
and some sharp maxima and minima appear for both transmitted
beams. In fact, by changing the incident angle the acquired phase
($k_{y}^{\prime}d$) of every propagating wave along the barrier
region varies, which leads to the oscillation of transmission
probabilities. The peak positions of the two beams are almost the
same. The corresponding GH shifts (red lines) of the two beams
exhibit some strong peaks beside usual ones with positive and
negative values. In order to explain qualitatively the behavior of
lateral shifts and the occurrence of their peaks, we rewrite the
formula (13) and (14) \cite{Article} as
\begin{equation}\label{15}
    \Delta_{tr}=\frac{\frac{d}{dk_{x}}\tan\varphi_{t}}{1+\tan^{2}\varphi_{t}}\ ,
\end{equation}
where
$\tan\varphi_{t}=\mathrm{Im}[t(k_{x})]/\mathrm{Re}[t(k_{x})]$. We
approximate $\mathrm{Re}[t(k_{x})]$ and $\mathrm{Im}[t(k_{x})]$
around a given point $k_{x_{0}}$ by retaining the first and second terms
of their Taylor expansion as $\mathrm{Re}[t(k_{x})]\simeq a_{R}+
b_{R}(k_{x}-k_{x_{0}})$ and $\mathrm{Im}[t(k_{x})]\simeq a_{I}+
b_{I}(k_{x}-k_{x_{0}})$, where $a_{R}, b_{R},a_{I}$ and $b_{I}$
are coefficients of the expansions. By inserting these
approximations in Eq. (15) we obtain
\begin{equation}\label{16}
    \Delta_{tr}=\frac{a_{R}b_{I}-a_{I}b_{R}}{|t(k_{x})|^{2}}.
\end{equation}
Moreover, by approximating
$\frac{d}{dk_{x}}|t(k_{x})|\simeq\frac{a_{R}b_{R}+a_{I}b_{I}}{|t(k_{x})|}$,
 Eq. (16) can be written as
\begin{equation}\label{17}
   \Delta_{tr}=\frac{a_{R}b_{I}-a_{I}b_{R}}{(a_{R}b_{R}+a_{I}b_{I})^2}(\frac{d}{dk_{x}}|t(k_{x})|)^{2}.
\end{equation}

From Eqs. (16) and (17), the local properties of $\Delta_{tr}$ in
the vicinity of a given point $k_{x_{0}}$, and hence, $\theta_{0}$
can be studied. According to Eq. (17), the absolute value of
$\Delta_{tr}$ at any point depends on the absolute value of the
slope of transmission probability at that point. By approaching
the sharp maxima and minima points in the blue curves, the slope
of the $T_{1,2}$ rapidly finds very large values. Therefore, the
absolute values of the corresponding GH shifts near these points
in the red curves suddenly increase, creating sharp maxima and
minima. The sign of GH shift is determined by the sign of the
numerator in Eq. (17). Some deep minima (not exactly zero) for
$T_{1}$ and $T_{2}$ appear in Figs. 6(a) and (b), specially for
$T_{2}$, i.e. $|t(k_{x})|\approx0$. According to Eq. (16), the absolute value of the
corresponding lateral shifts at these points can become large
and therefore, local maxima appear at these
points, as seen in red curves. Spatial separation between the two
beams and the angle between their spin orientations as a function
of incident angle are shown in Fig. 6(c) with blue and red curves,
respectively. One can see that at the given angle window, $\delta\Delta_{tr}$ exhibits 
several pronounced positive peaks, which make
the observation of well-separated beams practically more feasible.
Note that although at these points the transmission probabilities 
are far from the perfect splitter case with
$T_{1}=T_{2}=0.5$, these values are practically considerable. As
an example, for the incident angle $\theta=76.3^{\circ}$ at which
the obtained transmission probabilities for the two beams are
$0.2$ and $0.11$ (see Figs. 6(a) and (b)), the spatial separation
is about $6.5 \mu m$ which is large enough to measure
experimentally. At this point, the angle between spin orientations
of the two beams is $77.8^{\circ}$.
\begin{figure}[ht]
\begin{center}
\includegraphics[scale = 0.3]{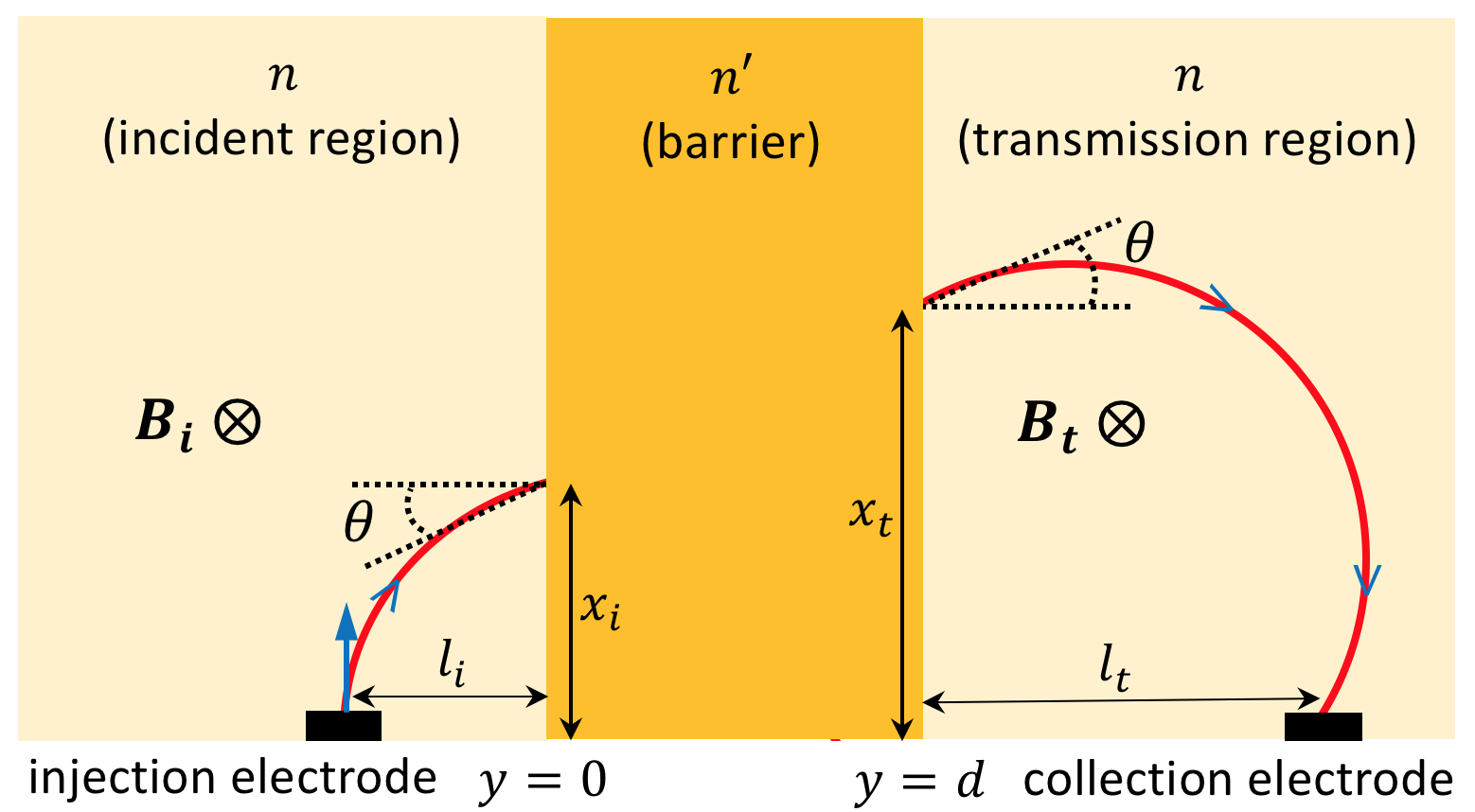}
\caption{(Color online) Schematic view showing the procedure of GH
shift measurement. The cyclotron radius $r_{i}$ in incident region,
fixed by both transverse magnetic field $\mathbf{B_{i}}$ and the Fermi
energy, determines the incident angle $\theta$ and also the impact
point of electron trajectory at the interface $y=0$. A varying
transverse magnetic field $\mathbf{B_{t}}$  induces a resonant
path with a specific radius $r_{t}$ which determines the entry
point of electron at the interface $y=d$. }\label{F8}
\end{center}
\end{figure}

 The transmission probabilities of the two transmitted beams and the
 corresponding GH shifts in terms of barrier width $d$ are depicted
 in Figs. 7(a) and (b). By varying the width of the barrier, the acquired phase,
$k_{y}^{\prime}d$, changes, and hence, $T_{1}$ and $T_{2}$
 oscillate by revealing several
sharp maxima and deep minima. We note that $k_{y}^{\prime}$ does
not change at a fixed incident angle. The behavior of GH shifts in
Figs. 7(a) and (b), compared to their corresponding transmission
probabilities, are similar to the behavior of GH shifts in Fig.
6(a) and (b), compared to their corresponding transmissions. That
means, near the sharp maxima of transmission probabilities where
their slope rapidly increases  with $d$, the absolute of the
corresponding GH shifts increases as well. Also, near the deep
minima of transmission probabilities where $|t(d)|\approx0$, the
absolute of the corresponding lateral shifts can be large.
Such a similarity is explained as follows:
The dependence of transmission coefficients, and hence, their
phases on $d$ is through the exponential function
$e^{ik_{y,n}^{\prime}d}$, when the boundary conditions of
continuity of the wave function in Eq. (4) and its derivative at
the interface $y=d$ are applied. Consequently, the dependence of
GH shifts on $d$ should be through exponential functions
$e^{ik_{y,n}^{\prime}d}$ as well. On the other hand, the
dependence of transmission coefficients as well as GH shifts on
$k_{x}(\theta)$ comes from $e^{ik_{y,n}^{\prime}d}$ and also from other terms which
vary slowly compared to the exponential functions. Therefore, when $d$ changes, the dependence of GH shift
on the transmission probability will be similar to the dependence
of GH shift on the transmission probability, when $k_{x}$ or
$\theta$ changes.

Spatial separation between the two electron beams as a function of 
$d$ is shown in Fig. 7(c). One can see that at $d\sim775A^{\circ}$, the separation between
the two beams is almost $10\mu m$ and the transmission probabilities of $T_{1}$ and 
$T_{2}$ are $\sim0.5$ and $0.2$, respectively (see Figs. 7(a) and (b)).
Also, the absolute value of $\delta\Delta_{tr}$ peaks between 720 \AA\,and 
800 \AA\,exhibits a considerable width, similar to $T_{1}$ and $T_{2}$. 
Moreover, the angle between spin orientations of two beams is
$59.1^{\circ}$ which is independent of $d$ and can be obtained
by the values of $E_{F}$ and $\theta$. Therefore, our findings
reveal that TIs with hexagonal warping effects can be utilized to
design an electron beam splitter with the ability of spatial
separation as large as a few micrometers with high chance of
observation of the well-separated beams.

In this paper we focused on a barrier with interfaces along $x$ direction ($\Gamma$K direction in 
${\bf k}$-space) which resulted in double refraction and double GH shifts of electron beams. If the barrier 
extends along $y$ direction ($\Gamma$M direction), due to the highly anisotropic nature of
hexagonally warped Fermi surface, triple refraction \cite{An-2012,M. Wu} and consequently triple GH shifts can
emerge. The occurrence of double and triple GH shifts in different directions can be a signature
of hexagonally warped Fermi surface, while they do not occur in other cases such as trigonally or tetragonally
warped Fermi surfaces. Also, observing gaps in the GH shift measurements in terms of electron energy 
can indicate the existence of energy gap in the band structure of matterials \cite{Ghadiri1}. Nevertheless, 
determining whether the shape of the Fermi surface can be identified with GH shift measurements, 
requires more research. It is worth mentioning that in Weyl semimetals it has been shown that the
GH and IF shifts of a reflected beam from a gapped medium can provide a probe of the topological
Fermi arc at the reflecting surface \cite{Chattopadhyay}.

\section{A proposal for GH shift measurement}

To the best of our knowledge the GH shift in electronic systems
has not been experimentally measured yet, due to the smallness of
GH shift values and the difficulty in producing a well collimated
electron beam \cite{Chen4,Yu3}. Although the magnitude of GH shift
in total reflection from a single-interface (step potential) is
about Fermi wavelength of electron which impedes its direct
measurement, it can be enlarged by considering a system acting as
a waveguide which causes accumulation of shifts in multiple
reflection of electron beam from the waveguide boundaries
\cite{Beenakker,Sun,Ghadiri1,Yu3}. Also, in the process of
transmitting electrons through potential barrier/well,
transmission resonances can occur which enhance the GH shift value
considerably \cite{Chen2,Song,Ghadiri,Ghadiri1,Zheng,Article1}. 
Note that similar and other mechanisms for amplifying optical GH 
shifts are considered in literatures (see Ref. [\onlinecite{Li2004}] and 
[\onlinecite{ChenAPL2017}]).
On the other hand, to directly measure GH shift values, we need a
collimator to generate collimated electron beam and then detect
the transmitted/reflected  beam from the interface by local gates.
Although various proposals for electron collimation in
2D materials \cite{Park,Cheianov,Moghaddam,M.
Liu,Betancur} and surface states of TIs \cite{Hassler,M. Liu} are
suggested, a decent production of narrow and well collimated
electron beam in such materials has not been attained yet \cite{S.
Chen,Libisch,Zhou,LaGasse}.

Despite the lack of efficient collimation, Chen et al. \cite{S.
Chen} achieved a direct measurement of angle-dependent
transmission probability based on TMF measurement scheme. They
applied a transverse magnetic field on electrostatically defined
$n$-$n'$ (p-n) junction on graphene and measured the
transresistance proportional to the transmission of electrons
between an injection electrode (at $n$ side) and a collection
electrode (at $n'$ side), while sweeping magnetic field and gate
voltage of $n'$ side. In this way, they reached a map in which the
first and higher-order resonant peaks appeared. Moreover, using a
semiclassical Billiard model they performed a simulation of
electron trajectories whose result was well-matched with that of
experimental data. As a result, they reverse-engineered the
first-order resonant transport by considering a trajectory for
electrons similar to the one that we consider in Fig. 8 which
clearly gives the peak positions observed in the experiment as
well as in the simulation.

Here, by applying a similar TMF measurement scheme we propose a
procedure for electrons' GH shift measurement on the surface of a
TI junction, as schematically shown in Fig. 8. We consider a positive 
GH shift which mostly occur in n-n$'$-n case. 
Before explaining the procedure, we give a brief discussion about 
survivability of surface states in the presence of a transverse magnetic field. 
In TMF phenomenon, it is assumed that the motion of electrons is ballistic, 
following the classical trajectories \cite{Taychatanapat,Milovanovic,Tsoi}. This is justified when the electron mean 
free path $l_{e}$ is larger than the width of the device in $x$ direction as well as the 
separation between the electrodes and interfaces $(l_{i}$ and $l_{t})$. The length of $l_{e}$ is 
estimated 120 nm for surface electrons in Bi$_{2}$Te$_{3}$ \cite{Xie-2017}. When the surface classical 
electrons are subjected to a transverse magnetic field $B$, they follow circular cyclotron orbits 
with radius $r=E_F/(ev_FB)$ given by Lorentz force, where $e$ is the charge of electron. 
If the system is treated quantum mechanically, these orbitals get quantized into Landau levels 
giving rise to chiral edge states. However when the magnetic field is not too high, the Landau 
levels undergo a collapse transition and the edge states can be avoided \cite{Akzyanov,Li-2015}.
To match our procedure with the above-mentioned experiment \cite{S. Chen}, we consider
electrons with low Fermi energy in the incident and transmission
regions with a circular shape of energy contour similar to that of
graphene, which makes an accurate control of electron trajectories
in the presence of magnetic field \cite{Yu}. In the barrier ($n'$)
region, there is no magnetic field and the warping effect can be
remarkable for large enough $V_{0}$ values. Under a transverse
magnetic field $B_{i}$, applied on incident electrons injected
from a narrow injection electrode, located at the bottom of this
region, the electrons undergo a cyclotron motion with radius 
$r_{i}=\frac{E_{F}}{e\upsilon_{F}B_{i}}$. With some simple
calculations, the impact point of electrons and their incident
angle at the interface $y=0$ can be determined as
$x_{i}=\sqrt{l_{i}(2r_{i}-l_{i})}$ and
$\theta=\arctan(\frac{r_{i}-l_{i}}{x_{i}})$, respectively, where
$l_{i}$ is the distance between the injection electrode and the
interface (see Fig. 8).

Now we apply an independent transverse magnetic field $B_{t}$ on
the transmission region. $B_{t}$ bends the transmitted electrons' path
downward the device into a cyclotron orbit. If the electrons enter
the collection electrode placed at the bottom of this region, a
peak in the transresistance between the injection and collection
electrodes (or corresponding voltage) will occur. Therefore, by
tuning $B_{t}$ in the experiment, it is possible to obtain the
amount of $B_{t}$ and the corresponding radius
$r_{t}=\frac{E_{F}}{e\upsilon_{F}B_{t}}$ for which a resonance in 
magnetic focusing takes place. Having $r_{t}$ and knowing the angle of
incoming electrons (equal to the incident angle) into this region
and the distance $l_{t}$ between the collection electrode and the
interface $y=d$, after some straightforward algebraic
computations, one can determine the position of entry point of
electrons at the interface $y=d$ as
$x_{t}=r_{t}\cos\theta+\sqrt{r_{t}^{2}-(l_{t}-r_{t}\sin\theta)^{2}}$. Finally
the GH shift can be obtained by $\Delta_{tr}=x_{t}-x_{i}$.

It is important to note that there is a correspondence between the variables 
used in Chen et al. experiment \cite{S. Chen} and the variables in our proposal. In their 
experiment, the same magnetic field $B$ is applied to both incident and transmission 
regions, and the magnetic field $B$ as well as the gate voltage of transmission region 
are varied. Also, the angle of entry of electrons into the transmission region $(\theta')$ is 
different from the incident angle $\theta$. Moreover, $\theta'$ is a function of $\theta$ 
according to the Snell’s law $\sin\theta'=((E_F-V_i)/(E_F-V_t ))\sin\theta$, where $V_{i}$ 
and $V_{t}$ are the gate voltage of incident and transmission regions, respectively. In our 
proposal, different magnetic fields $B_{i}$ and $B_{t}$ are applied to the incident and 
transmission regions, respectively. The gate voltage of the barrier region is fixed, while 
$B_{i}$ and $B_{t}$ are variable during the experiment. On the other hand, the transmitted 
electrons enter the transmission region with the same angle as the incident angle $\theta$ 
but with a lateral shift $\Delta_{tr}$ which is a function of incident angle. Because of such 
correspondences, we expect that the electrons contributing in the transresistance peak to 
be those electrons that leave the injection electrode vertically, just like the 
Chen et al. experiment  \cite{S. Chen}. 

The above discussion can be presented in a general form as follows. Consider the electrons 
that leave the injection electrode with arbitrary injection angle $\beta$ (with respect to $y$ axis) 
at a given $B_{i}$. The incident angle of these electrons can be calculated as 
$\theta=\arcsin(\sin\beta-l_{i}/r_{i})$. At a fixed $B_{t}$, due to the dependence of 
impact point of electrons on the interface at $y=0$ and also $\Delta_{tr}$ to $\theta$ the electrons 
reach the edge of the device ($y$ axis) in the transmission region at a position that depends on 
their injection angle $\beta$. Calculating the derivative of $\theta$ with respect to $\beta$, we 
obtain $d\theta/d\beta=\cos\beta/\sqrt{\cos^{2}\beta+2(l_{i}/r_{i})\sin\beta-l_{i}^{2}/r_{i}^{2}}$ whose 
magnitude is zero at $\beta=\pi/2$. This means that the electrons which leave the injection 
electrode in small vicinity of angle $\beta=\pi/2$, have the same incident angle $\theta$, and hence, 
the same lateral shift $\Delta_{tr}$, resulting the largest density of electrons at a point on the edge 
of the device in the transmission region. By sweeping the magnetic field $B_{t}$, $r_{t}$  is varied, 
so that at a fixed collection electrode position, a peak in transresistance belonging to the 
electrons that leave the injection electrode vertically, appears. By tuning $B_{t}$ on larger amounts, 
the cyclotron radius $r_t$ is reduced, so that the electrons can reach the collection electrode after one 
or more specular reflection from the interface and/or the edge of the device, leading to the formation 
of next peaks \cite{Tsoi}.

Although the present proposal of GH shift measurement was applied
to the surface state of Bi$_{2}$Te$_{3}$  consisting of a
single nondegenerate Dirac cone, this approach can also be
utilized in 2D conventional systems such as graphene and other single-layer hexagonal crystals.
Nevertheless, in materials consisting of multivalleys,
multirefraction can appear, making the observation of GH shift
more complicated than the present study. Moreover, surface states
of TIs are topologically protected against non-magnetic
perturbations compared to the conventional surface states which
are sensitively dependent on the geometry of surface structure.

Since in most Dirac materials the GH shift is spin and/or valley dependent which originates from spin-orbit 
coupling \cite{Sun, Ghadiri,Ghadiri1,Azarova,Article1,Article2}, 
the measurement of GH shift can provide the possibility of fabrication of 
spin/valley devices based on electronic beam shifts.

\section{Conclusion}

In summary, we studied theoretically the influence of hexagonal
warping effect on the transport properties and lateral shifts of 
electrons at the surface of a TI n-n${'}$-n (n-p-n)
junction. It is shown that double refractions occur when the Fermi
energy and incident angle of electron beams exceed their
critical values. We establish an expression for calculating GH
shift values and show that a deflection of propagation direction
of beams from their central propagation directions appears due to
the hexagonal warping effect. The dependence of lateral shifts and the
corresponding transmissions on system parameters such as incident
angle, height and the width of potential barrier are carefully
examined. We show that the system can produce two spatially
separated beams with different spin orientations as a result of GH
effect. Therefore our findings provide an alternative way to
construct an electron beam splitter on the basis of TI junctions. Using the physics of TMF phenomenon, we
also introduce a procedure for experimentally measuring the GH
shift of electron beams in 2D electronic systems which may pave a new route in
spin/valley-tronics.

\section{ACKNOWLEDGMENTS}
This work was supported by Iran National Science Foundation: INSF
(Grant No. 96017337).


\begin{thebibliography}{99}
\bibitem{Article3}
At $V_{0}=E_{F}$, the angle in which TIR begins is dependent on $d$ value. 
By increasing $d$, although the angle decreases, it never reaches zero due to the Klein tunneling effect.
\bibitem{Article}
In Eq. (14), the second and the third terms can be ignored in the vicinity of the peak positions, compared to the first term.
\bibitem{Article2}
In graphene, the electronic GH shift is spin and valley independent. 
However, with magnetic barrier and/or considering small intrinsic spin-orbit coupling or extrinsic Rashba-type spin-orbit coupling, the GH shift will depend on spin and valley.
\end{thebibliography}
\end{document}